\documentclass[a4paper,twoside]{article}

\usepackage{microtype}
\usepackage{framed}
\usepackage[utf8]{inputenc}
\usepackage{graphicx}
\usepackage{subfigure}
\usepackage{url}
\usepackage{calc}
\usepackage{amssymb}
\usepackage{amstext}
\usepackage{amsmath}
\usepackage{amsthm}
\usepackage{multicol}
\usepackage{pslatex}
\usepackage{natbib}
\usepackage{apalike}
\usepackage{SCITEPRESS}

\begin{document}

\title{Inferring causality from noisy time series data\subtitle{A test of Convergent Cross-Mapping} }

\author{\authorname{Dan Mønster\sup{1,2,*}, Riccardo Fusaroli\sup{2,3}, Kristian Tylén\sup{3,2}, Andreas Roepstorff\sup{2}, Jacob F. Sherson\sup{4,5}}
\affiliation{\sup{1}Department of Economics and Business Economics, Aarhus University, Fulglesangs All\'{e} 4, 8210 Aarhus V, Denmark}
\affiliation{\sup{2}Interacting Minds Centre, Aarhus University, Jens Chr. Skous Vej 4, 8000 Aarhus C, Denmark}
\affiliation{\sup{3}Center for Semiotics, Aarhus University, Jens Chr. Skous Vej 2, 8000 Aarhus C, Denmark}
\affiliation{\sup{4}Department of Physics and Astronomy, Aarhus University, Ny Munkegade 120, 8000 Aarhus C, Denmark}
\affiliation{\sup{5}AU Ideas Center for Community Driven Research, Aarhus University, Ny Munkegade 120, 8000 Aarhus C, Denmark}
\email{\sup{*}Corresponding author: danm@econ.au.dk}
}


\abstract{Convergent Cross-Mapping (CCM) has shown high potential to perform causal inference in the absence of models. We assess the strengths and weaknesses of the method by varying coupling strength and noise levels in coupled logistic maps. We find that CCM fails to infer accurate coupling strength and even causality direction in synchronized time-series and in the presence of intermediate coupling. We find that the presence of noise deterministically reduces the level of cross-mapping fidelity, while the convergence rate exhibits higher levels of robustness. Finally, we propose that controlled noise injections in intermediate-to-strongly coupled systems could enable more accurate causal inferences. Given the inherent noisy nature of real-world systems, our findings enable a more accurate evaluation of CCM applicability and advance suggestions on how to overcome its weaknesses.}

\onecolumn \maketitle \normalsize \vfill

\section{\uppercase{Introduction}}
\label{sec:introduction}
\noindent The ability to infer causality from the relation between two variables is an intensely researched area. Common approaches involve having both a model of the system being studied and a series of measurements of that same system \citep{pearl_causal_2009}. In many cases, however, we do not have an adequate model of the system, or face several conflicting models. Complex natural, technical and social systems are prime examples, including ecosystems, brains, the climate and the global financial system. In such cases, inferring whether one part has a causal influence on another part has to rely on model-free methods. Convergent Cross-Mapping (CCM) \citep{sugihara_detecting_2012} is a relatively new method that promises to `distinguish causality from correlation' in time series data (ibid.,~p.~496). CCM was introduced as an alternative to other methods that detect causality between two time series, principally Granger causality~\citep{granger_investigating_1969}. 

Granger causality has been developed to assess easily separable linear systems, whereas CCM is primarily suited for weakly coupled components of non-linear dynamic systems. Accordingly, they have slightly diverging notions of causality. Therefore, the statement `$X$ causes $Y$' would more accurately be phrased as `$X$ Granger causes $Y$' or `$X$ CCM causes $Y$'. \citet{sugihara_detecting_2012} refer to the type of causality captured by CCM as \emph{dynamic causation}, reminiscent of what \citet{lakoff_why_2010} terms systemic causation. For a system with several variables, for which time series data are available, the CCM method produces a causal network structure describing which variables are causally connected, including the direction of causality. Like mutual information \citep{fraser_independent_1986}, transfer entropy \citep{schreiber_measuring_2000} and cross-recurrence quantification \citep{webber_dynamical_1994,marwan_recurrence_2007} CCM is a state space method relying on time-delayed embedding of the time series data in a higher dimensional space.

CCM has already been used in a wide range of different fields for different kinds of data \citep{mcbride_sugihara_2015,bozorgmagham_causality_2015,mccracken_convergent_2014}, and it has been noted \citep{mccracken_convergent_2014} that CCM results are not always consistent with theoretical intuitions. \citet{mccracken_convergent_2014} extended CCM to pairwise asymmetric inference (PAI), which they demonstrated to give results that are in better accordance with the intuitively expected outcomes for several physical systems. Recently \citet{ma_detecting_2014} developed cross-map smoothness (CMS)---a method related to CCM---which has the advantage of requiring fewer points in the time series. Notice, however, that in this paper we focus on CCM, leaving the analysis of developments such as PAI and CMS for future work.

Given the interest and relevance of CCM, it is important to better understand its strengths and limitations. Accordingly, in this article we will present results of an in-depth study of a simple model system, the coupled logistic map, with particular emphasis on how the CCM results reflect the model input, varying strength of coupling, and levels of noise. We will briefly describe how CCM is applied to the model. Then we look at how the choice of coupling affects the CCM results, and finally we report on results of adding noise to the system. Although noise in real-world data is ubiquitous, the inclusion of noise in model investigations has been largely neglected. We find that noise can dramatically change the strength of causal inferences. Crucially, we also observe that  the appropriate injection of noise into the dynamics can be used as means of inferring the relative strengths of the coupling (to the extent that the system can be controlled). 

\section{\uppercase{Model system}}
\begin{figure}[!h]
  \vspace{-0.2cm}
  \centering
   {\includegraphics[width=\columnwidth]{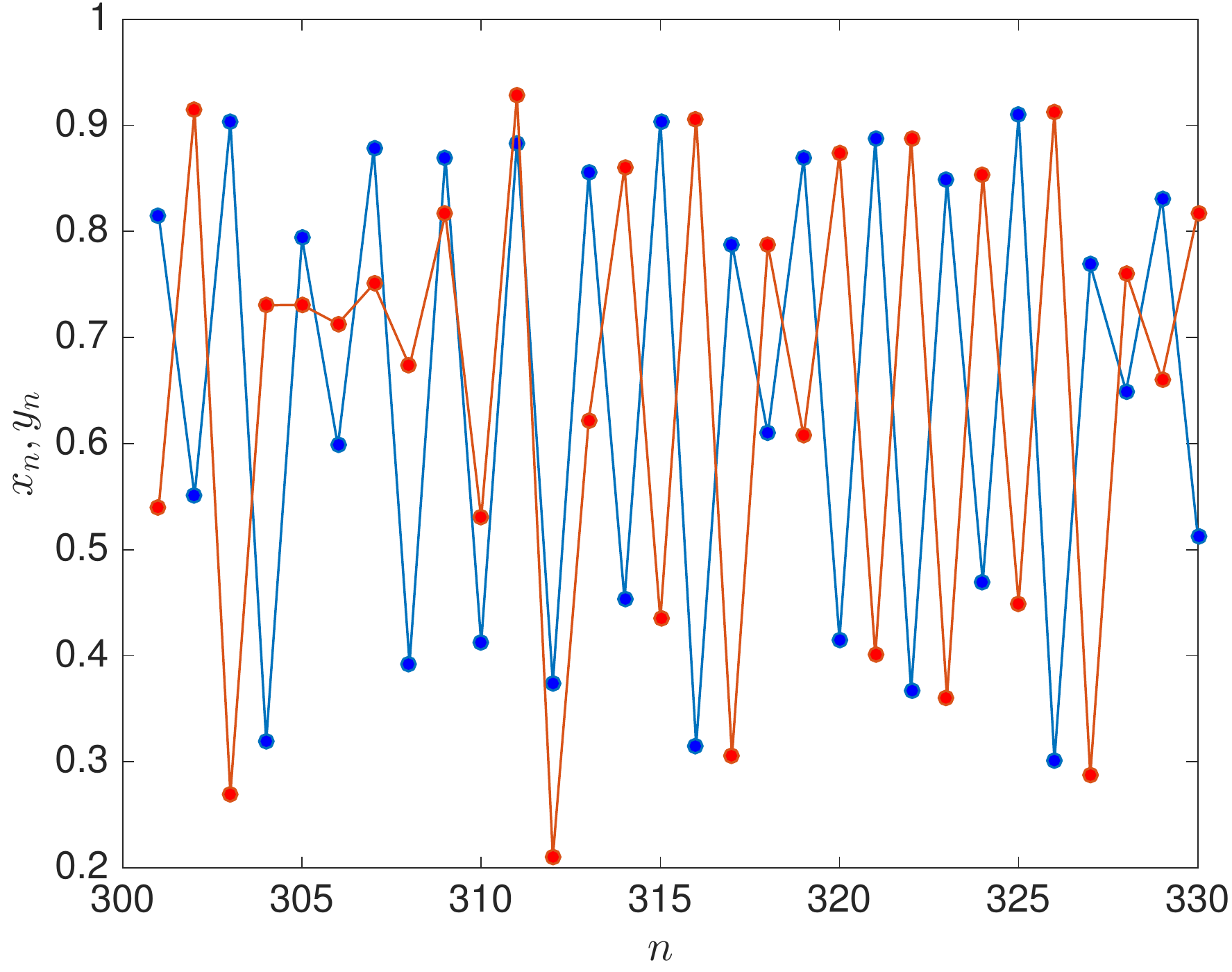}}
  \caption{Plot of the time series of $x$ (shown in blue) and $y$ (shown in red) after the initial 300 time steps. Note that $\beta_{xy}=0$, so there is no coupling from $y$ to $x$ and the causality is therefore unidirectional from $x$ to $y$ $(\beta_{yx}=0.05)$. Both $r_x=3.65$ and $r_y=3.77$ are chosen to be in a chaotic regime of the logistic map.}
  \label{fig:timeseries}
  \vspace{-0.1cm}
\end{figure}

\noindent The logistic map has long been a model system of nonlinear dynamics displaying regular periodic behavior as well as deterministic chaos \citep{may_simple_1976}. A system of two coupled logistic maps has been used as a simple model of chemical reaction dynamics \citep{ferretti_study_1988} and population dynamics \citep{lloyd_coupled_1995}. This makes it an ideal model system for testing CCM. We follow \citet{sugihara_detecting_2012} and study two logistic maps coupled through linear terms
\begin{equation}\label{eq:model}
\begin{split}
    x_{t+1} = x_t(r_x(1-x_t)-\beta_{xy}y_t)\\
    y_{t+1} = y_t(r_y(1-y_t)-\beta_{yx}x_t)
\end{split}
\end{equation}
The two variables $x$ and $y$ have a nonlinear dependence on their own past values parameterized by the growth rates $r_x, r_y$, and are coupled to each other through linear terms with coupling constants that parameterize the strength of the coupling from $x$ to $y$ $(\beta_{yx})$ and from $y$ to $x$ $(\beta_{xy})$. Times series for $x$ and $y$ are shown in Figure~\ref{fig:timeseries} for a particular choice of the parameters in the model.

\begin{figure}[!h]
  \vspace{-0.2cm}
  \centering
   {\includegraphics[width=\columnwidth]{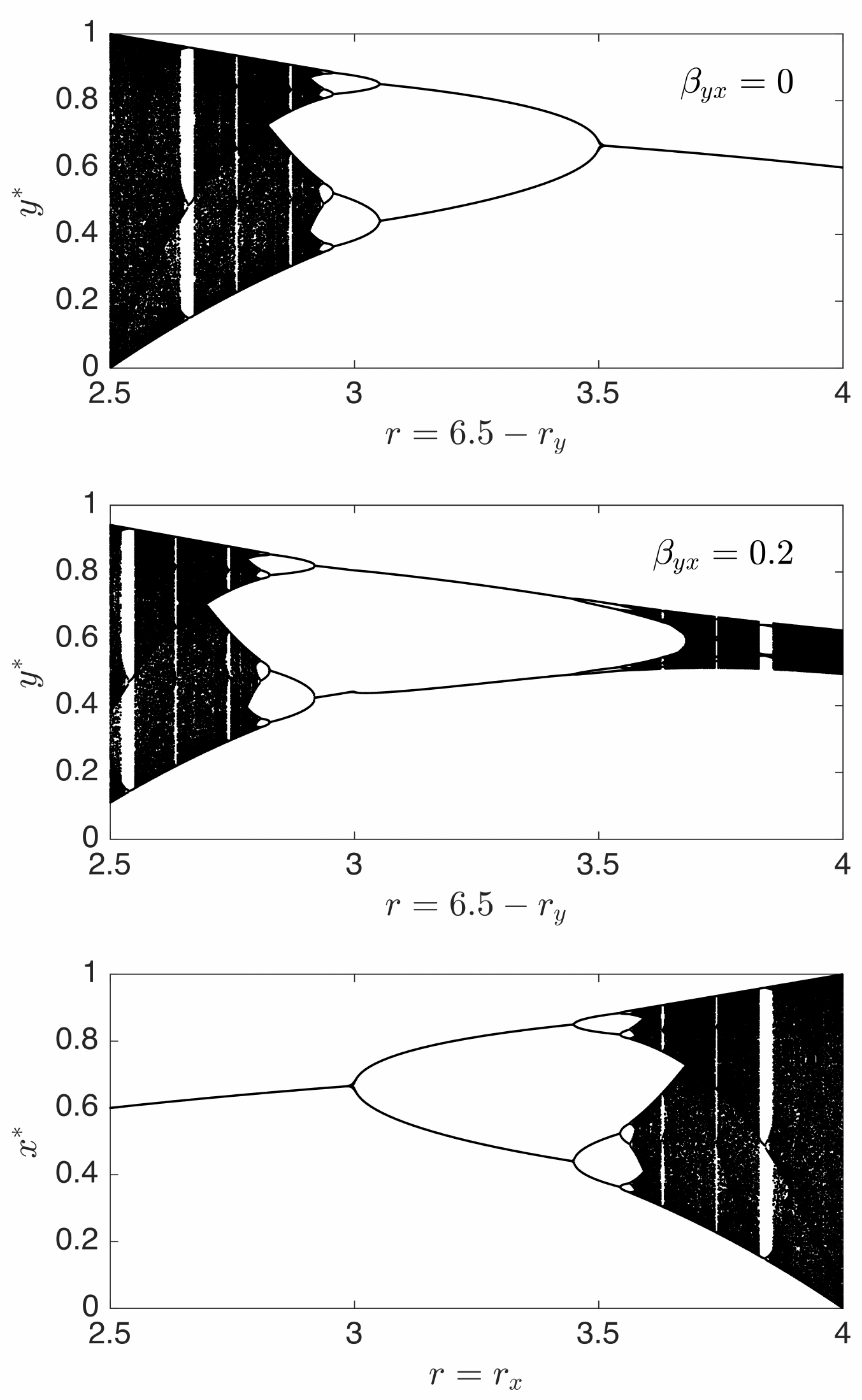}}
  \caption{Phase diagram for the coupled logistic map in Eq.~\ref{eq:model} for $\beta_{xy}=0$. The growth rates $r_x,r_y$ are parameterized by the variable $r$ on the horizontal axis as $r_x=r$ and $r_y=6.5-r$. For each value of $r$ the plots represent a different pair of parameters $r_x,r_y$. The bottom panel shows the fixed points $x^{*}$ as a function of the growth rate, and the top panel $(\beta_{yx}=0)$ shows the fixed points $y^{*}$. The center panel shows the fixed points for $\beta_{yx}=0.2$.}
  \label{fig:phase}
  \vspace{-0.1cm}
\end{figure}

A single logistic map has well-known domains of periodic behavior controlled by the growth rate $r$. As $r$ becomes larger, increasingly frequent period-doublings occur, which ultimately give way to chaotic behavior at $r\approx 3.57$. This well-known phase diagram or bifurcation diagram is shown in the bottom panel of Figure~\ref{fig:phase}, representing the $x$ variable. The top panel shows the phase diagram for another logistic map representing the $y$ variable. In order to illustrate the effect of coupling between two logistic maps the growth rate for the $y$ variable $r_y$ is related to $r_x$ by the equation $r_x+r_y = 6.5$. This is the equation for a line in the $(r_x,r_y)$-plane with a slope of $-1$. The middle panel in Figure~\ref{fig:phase} illustrates the phase diagram of $y$ for a family of two coupled logistic maps along this line in the $(r_x,r_y)$-plane as described by Eq.~\ref{eq:model} with $\beta_{xy}=0$ and $\beta_{yx}=0.2$.

The middle and lower panels show the effect of unidirectional coupling between two logistic maps. It is evident that the dynamics of $x$ leaves an imprint on $y$ as indicated by the fixed points. For low $r_y$ (high $r$) where the isolated system (top panel) is stationary, the effect of coupling from the chaotic dynamics of $x$ induces chaotic dynamics in $y$, albeit with a much smaller amplitude. For high $r_y$ (low $r$) where the isolated system is chaotic, the coupling from $x$ which is in the stationary domain has the effect of stabilizing the periodic phases of $y$ which persist to higher values of $r_y$ than in the isolated system.

\subsection{Noise}
Real-world systems present varying degrees of environmental and measurement noise and the application of the method to such systems thus critically depends on the method’s ability to cope with noise. We therefore also consider the model with added noise terms:
\begin{equation}\label{eq:noise}
\begin{split}
    x_{t+1} = x_t(r_x(1-x_t)-\beta_{xy}y_t) + \epsilon_{x,t}\\
    y_{t+1} = y_t(r_y(1-y_t)-\beta_{yx}x_t) + \epsilon_{y,t}
\end{split}
\end{equation}

Here $\epsilon_{x,t}$ and $\epsilon_{y,t}$ are the noise terms on $x$ and $y$ which we model as stochastic variables sampled from normal distributions $\mathcal{N}(0,\sigma^2)$ with zero mean and standard deviation $\sigma$. We will refer to $\sigma$ as the noise level. Note that noise in either variable in Eq.~\ref{eq:noise} propagates to later values of the variable and to the other variable through the coupling terms. Hence, the noise terms in this model represent perturbations from the environment rather than noise from the measurement process.

\section{\uppercase{Method}}
\begin{figure}[!h]
  \vspace{-0.2cm}
  \centering
   {\includegraphics[width=\columnwidth]{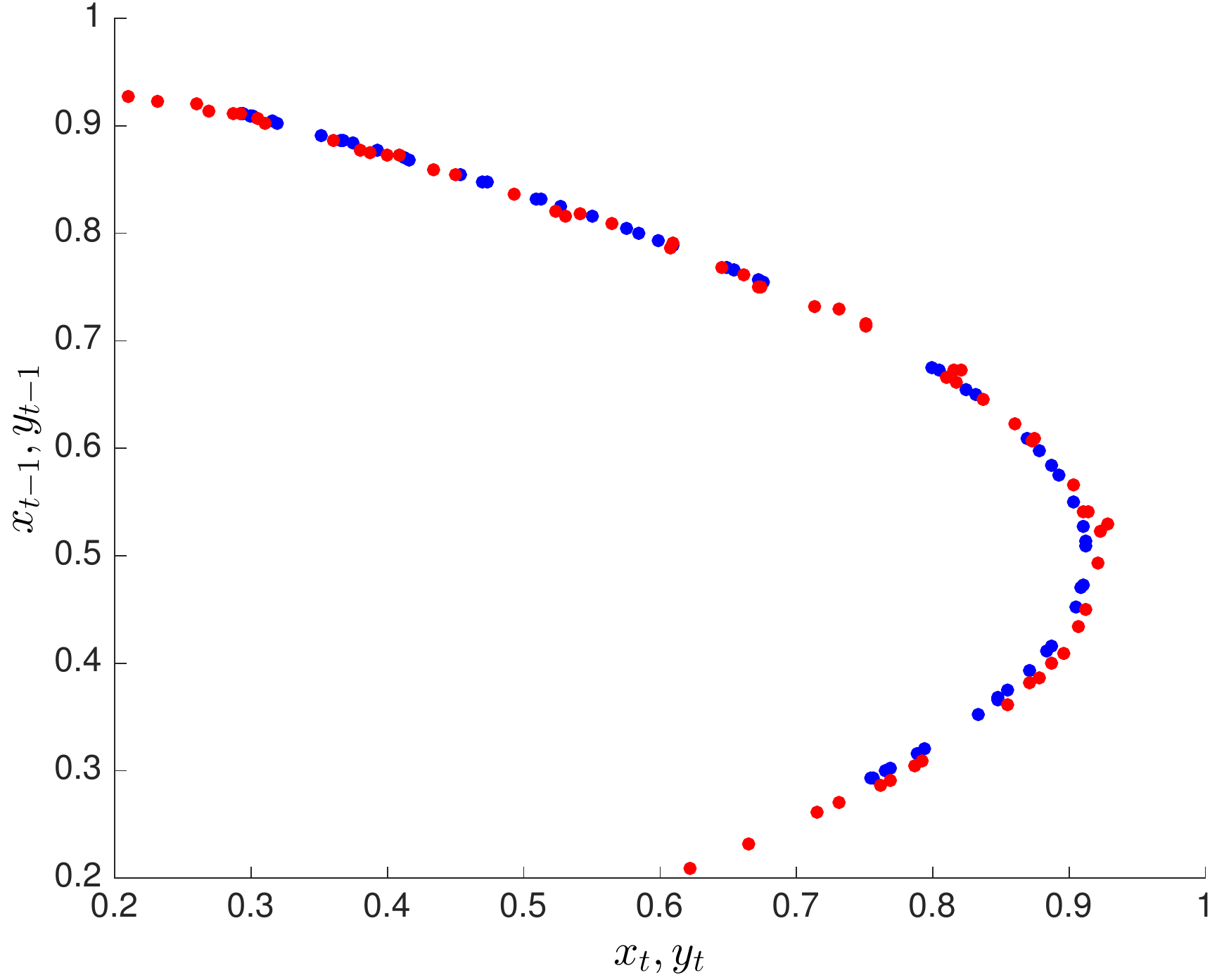}}
  \caption{The shadow manifolds based on the time series displayed in Figure~\ref{fig:timeseries}, embedded using $E=2$ and $\tau=1$. A total of 60 points $\mathbf{x}_t = (x_t,x_{t-1})$ in $M_x$ are shown in blue and the points $\mathbf{y}_t = (y_t,y_{t-1})$ in $M_y$ are shown in red.} 
  \label{fig:shadow}
  \vspace{-0.1cm}
\end{figure}

\noindent Convergent cross-mapping is a state space method that relies on Takens' theorem \citep{takens_detecting_1981} to reconstruct the underlying dynamics of a system in a model-free fashion, by using time-delayed embedding to reconstruct its attractor landscape (see, e.g., \citet{abarbanel_analysis_1996}). Our model system in Eq.~\ref{eq:model} can be described by its attractor, that is, the trajectory consisting of consecutive points in two-dimensional Euclidean space given by the Cartesian coordinates $(x_0,y_0), (x_1,y_1), \ldots ,(x_N,y_N)$ resulting from the dynamics described by Eq.~\ref{eq:model}. We discard the first 300 points to avoid transient behavior of the model from affecting our results. According to Takens' theorem we can approximately reconstruct the attractor from one of the variables $x$ or $y$ alone using time-delayed embedding of the points in one of the time series, say $x$, where each point in $E$-dimensional space is given by $\mathbf{x}_t  = (x_t, x_{t-\tau}, x_{t-2\tau}, \ldots x_{t-(E-1)\tau})$. The embedding thus depends on two parameters: the time delay $\tau$ and the dimension $E$ of the space in which the reconstructed attractor is embedded. Since we know that there are only two independent variables in our model system we choose $E=2$, but estimating the embedding dimension from the times series obtained from Eq.~\ref{eq:model} using the false nearest neighbor method \citep{abarbanel_analysis_1996} gives the same result. The time delay for the embedding was set to $\tau=1$ based on the average mutual information criterion \citep{kantz_nonlinear_1997}. The reconstructed attractors, referred to as \emph{shadow manifolds} by \citet{sugihara_detecting_2012}, are shown in Figure~\ref{fig:shadow}.

The two key ingredients of the CCM method are the concept of \emph{cross-mapping} and the \emph{convergence} property that are explained in the following sections.

\subsection{Cross-mapping}
The time series data from each variable---in our case $x$ and $y$---can be used to construct shadow manifolds---$M_x$ and $M_y$---that are approximations to the true attractor. Given the non-separability of the system, Taken's theorem demonstrates that when two different variables represent different parts of the same dynamical system their shadow manifolds are diffeomorphic to the true attractor and therefore to each other. Intuitively the two variables are connected because they are part of the same dynamical system as evidenced by the fact that they both represent a dimension in the state space. So, if $x$ has a causal influence on the dynamics of $y$ then $x$ will influence the dynamics of $y$. This `imprint' of $x$ on $y$ means that knowledge of the shadow manifold $M_y$ obtained from the time series of $y$ can be used to estimate values of $x$. This estimate is called the cross-map and is denoted $\hat{x}\vert M_y$.

To find the cross-mapped estimate $\hat{x}_t\vert M_y$ of $x_t$ we start by identifying the corresponding point $\mathbf{y}_t$ in $M_y$. Since $M_y$ is  diffeomorphic to $M_x$, a small region around $\mathbf{y}_t$ will map to a small region around $\mathbf{x}_t$ and this can be used to estimate $x_t$. To form a bounding simplex around $\mathbf{y}_t$ at least $E+1$ points are needed \citep{sugihara_nonlinear_1990}, so the $E+1$ nearest neighbors of $\mathbf{y}_t$ in $M_y$ are found. Sorted from the closest to the farthest point from $\mathbf{y}_t$ we call these $\mathbf{y}_{t_1}, \mathbf{y}_{t_2},\ldots , \mathbf{y}_{t_{E+1}}$. We use the points in the time series for $x$ at the corresponding times, i.e., $x_{t_1}, x_{t_2},\ldots , x_{t_{E+1}}$ to estimate $x_t$ as
\begin{equation}\label{eq:xmap}
	\hat{x}_t\vert M_y = \sum_{i=1}^{E+1} w_i x_{t_i}
\end{equation}
The weights $w_i$ are exponentially weighted with the Euclidean distance between $\mathbf{y}_t$ and the nearest neighbor points:
\begin{equation}\label{eq:weights}
	w_i = u_i \biggm/ \sum_{j=1}^{E+1} u_j, 
	\quad u_i = \exp\left( -\frac{\left\lVert \mathbf{y}_t-\mathbf{y}_{t_i} \right\rVert}{\left\lVert \mathbf{y}_t-\mathbf{y}_{t_1} \right\rVert} \right)
\end{equation}
where $\left\lVert \cdot \right\rVert$ is the Euclidean norm in $\mathbb{R}^E$.

\begin{figure}[!h]
  \vspace{-0.2cm}
  \centering
   {\includegraphics[width=\columnwidth]{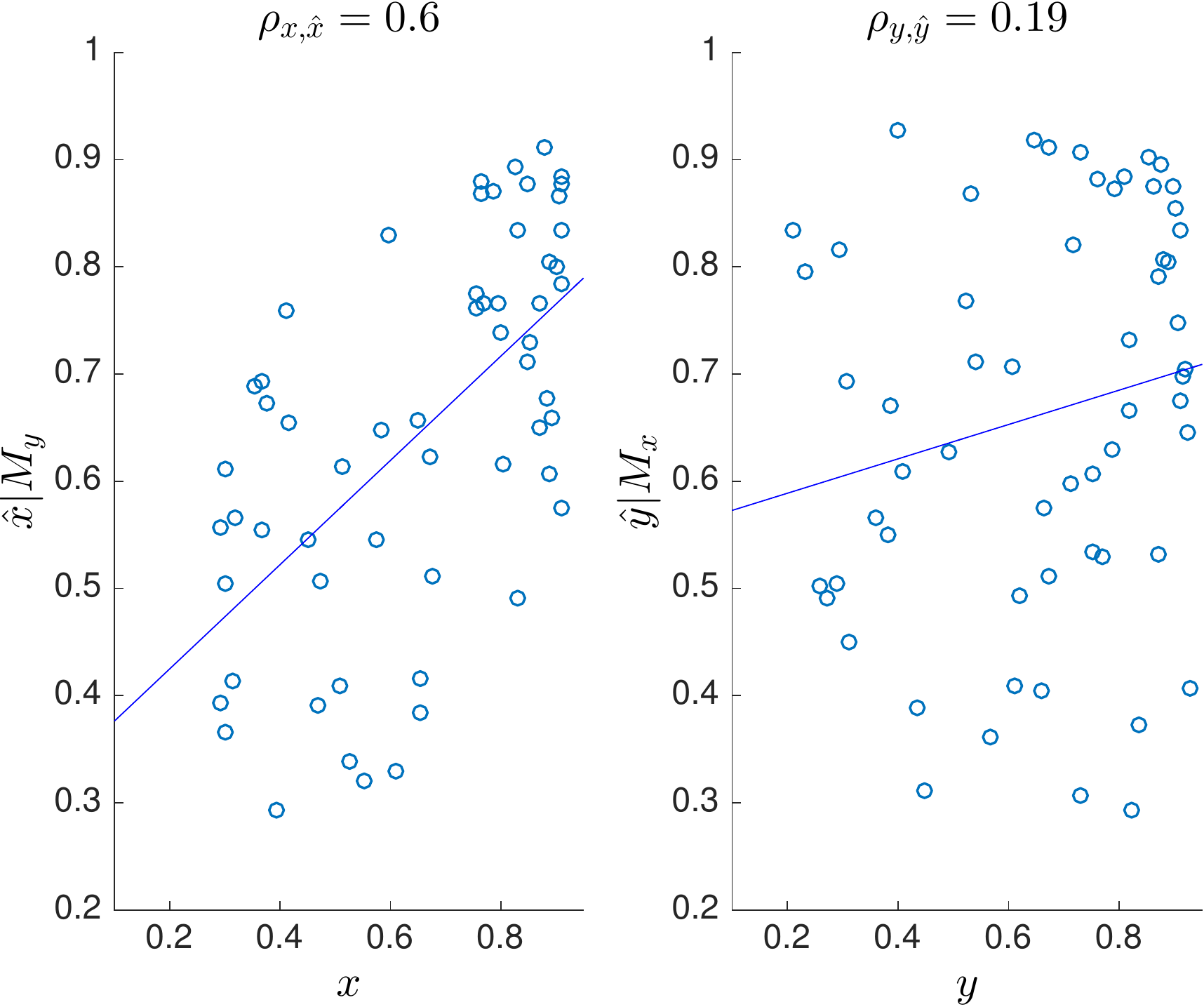}}
  \caption{Scatter plot of pairs of observed values of $x$ and estimated values $\hat{x}\vert M_y$ (left panel) and the equivalent for $y$ (right panel). The cross-mapped estimates were computed according to Eq.~\ref{eq:xmap} and were based on the shadow manifolds in Figure~\ref{fig:shadow}.} 
  \label{fig:scatter}
  \vspace{-0.1cm}
\end{figure}

Estimating one point alone is not sufficient to show how well $\hat{x}_t\vert M_y$ estimates the true value $x_t$. A library consisting of $L$ points from $M_y$ is therefore used to provide estimates of $L$ points in the time series for $x$. The Pearson correlation coefficient $\rho_{x\hat{x}}$ between the $L$ true values from $x$ and the $L$ cross-mapped estimates is an indicator of how much the dynamics of $x$ influences the dynamics of $y$. Scatter plots of observed values and cross-mapped estimates of $x$ and $y$ are shown in Figure~\ref{fig:scatter}. 

%
%

The results indicate that the cross-mapped estimates of $x$ obtained from $M_y$ $(\rho_{x\hat{x}}\approx 0.6)$ are better than the cross-mapped estimates of $y$ obtained from $M_x$ $(\rho_{y\hat{y}}\approx 0.2)$. It is therefore tempting to conclude that  $x$ causes $y$, but, although correct in this case, such a conclusion is potentially misleading. Indeed, A slight variation in the parameters, can produce results that would lead to the opposite conclusion. The additional criterion of convergence is crucial to correctly infer the direction and relative strength of causality.

Cross-mappings were performed with \textsc{xmap} \citep{monster_xmap_2013} which was developed in MATLAB and validated by reproducing the results by \citet{sugihara_detecting_2012} and by comparing with an independently developed algorithm \citep{jespersen_personal_2013}. A small regularizing term $\delta$ is added to the denominator in the argument to the exponential function in Eq.~\ref{eq:weights} in order to avoid floating point overflow in cases where $\mathbf{y}_{t_1} = \mathbf{y}_t$.

\subsection{Convergence}
The convergence properties of the correlation between observed values and cross-mapped estimates as a function of the library length $L$ is the second key ingredient in CCM. If $x$ causes $y$ then the estimate of $x$ obtained from $M_y$ should improve as the number of points $L$ sampled from $M_y$ becomes larger, since the library of samples will become a better and better representation of the attractor, and the nearest neighbor points will be closer and closer to $\mathbf{y}_t$.

\begin{figure}[!h]
  \vspace{-0.2cm}
  \centering
   {\includegraphics[width=\columnwidth]{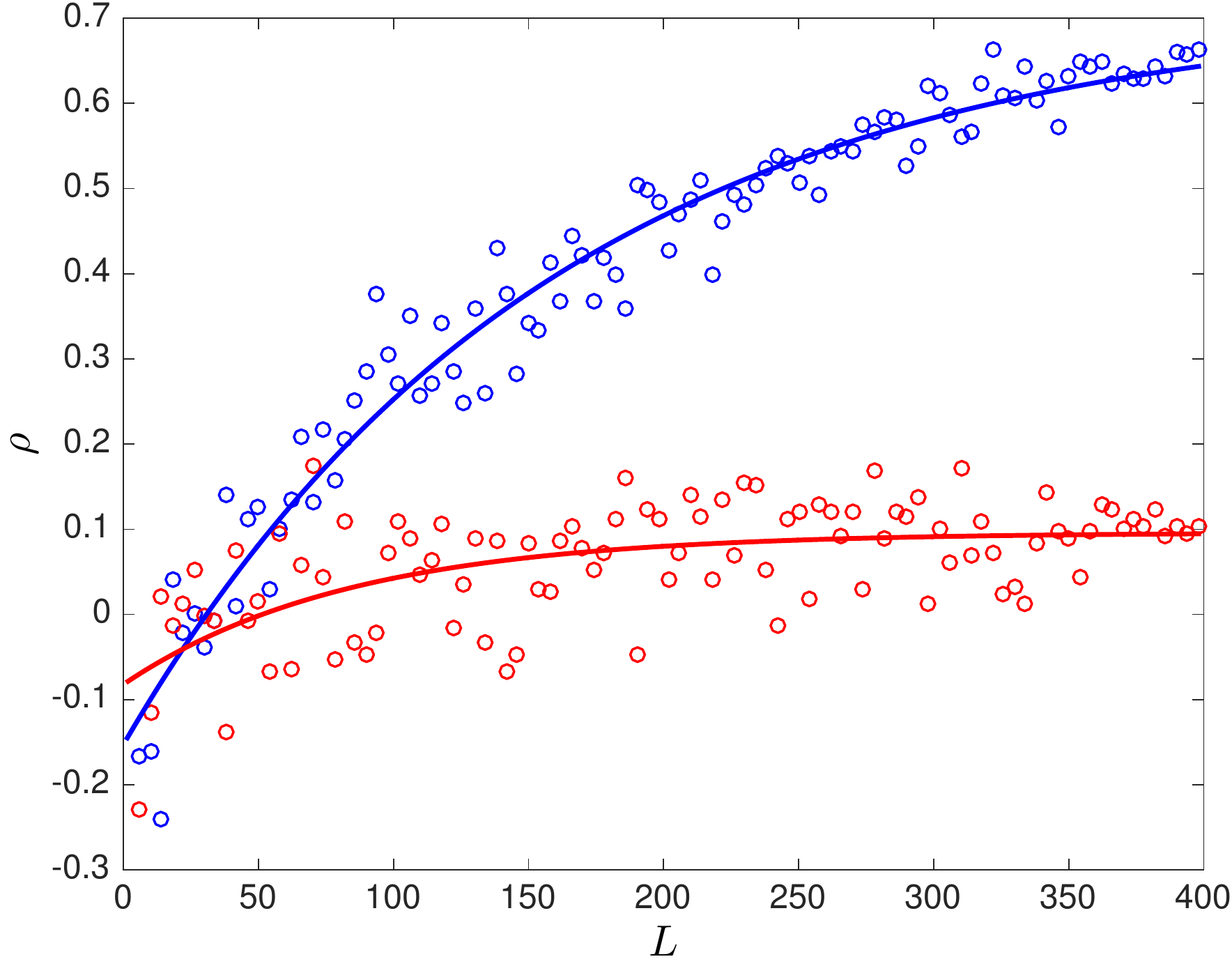}}
  \caption{The correlation coefficient $\rho_{x,\hat{x}}$ (blue circles) and $\rho_{y,\hat{y}}$ (red circles) as a function of library length $L$. Fits to the function in Eq.~\ref{eq:fit} are shown as solid lines. For $x$ the converged value is $\rho_\infty=0.71$, and for $y$ it is $\rho_\infty=0.096$. Parameter values: $r_x=3.65$, $r_y=3.77$, $\beta_{xy}=0$, $\beta_{yx}=0.05$.} 
  \label{fig:convergence}
  \vspace{-0.1cm}
\end{figure}

This convergence phenomenon is illustrated in Figure~\ref{fig:convergence} where correlations derived from the data shown in Figure~\ref{fig:scatter} have been computed for increasing values of the library length. Figure~\ref{fig:convergence} shows that $\rho_{x\hat{x}}$ converges toward a much higher value than $\rho_{y\hat{y}}$ does. This is evidence that $M_y$  enables much better estimates  for $x$ than $M_x$ does for $y$. Note that cross-mapped estimation skill is from $y$ to $x$ when $x$ causes $y$.

\subsubsection{Fitting the convergence}
Rather than relying only on the value of the correlation coefficient for the largest obtained value of $L$ we fit the computed values of $\rho$ as a function of $L$ to the following function
\begin{equation}\label{eq:fit}
\rho(L) = \alpha e^{-\gamma L} + \rho_\infty .
\end{equation}
This function, containing three constants $\alpha$, $\gamma$ and $\rho_\infty$, is in accordance with the computed values of the correlation coefficient. It can be interpreted as convergence toward the value $\rho_\infty$ as $L\to\infty$ where the speed of convergence is given by the constant $\gamma$. The third constant $\alpha$ is necessary to obtain a good fit, but does not provide additional interpretive value.

The solid lines in Figure~\ref{fig:convergence} represent the fitted correlation coefficients to the function in Eq.~\ref{eq:fit} using the same parameters as in Figure~\ref{fig:timeseries}. The fitted value for $\rho_{x\hat{x}}$ is  $\rho_\infty=0.71 \pm 0.046$ (95\% CI), and the fitted value for $\rho_{y\hat{y}}$ is $\rho_\infty=0.096 \pm 0.024$.

\section{\uppercase{Results}}
\noindent The previous section introduced CCM and demonstrated how to apply the method on the model system using a particular choice of parameter values as an example. \citet{sugihara_detecting_2012} summarize their results for different values of the coupling constants $\beta_{xy}$ and $\beta_{yx}$ with $r_x$ and $r_y$ chosen uniformly from the interval $[3.6,4]$ in Figure~3B of their manuscript. This figure indicates that CCM gets the direction of the coupling right, but also indicates that for some choices of the coupling constants the method may give results that are hard to distinguish from symmetric coupling, even when $\beta_{xy} \gg \beta_{yx}$ or vice versa. 

To investigate this further, we will look at what happens for a particular choice of $r_x$ and $r_y$ when the coupling between the two variables is manipulated.

\subsection{Effect of weak and strong coupling}
\begin{figure}[!t]
  \vspace{-0.2cm}
  \centering
   {\includegraphics[width=\columnwidth]{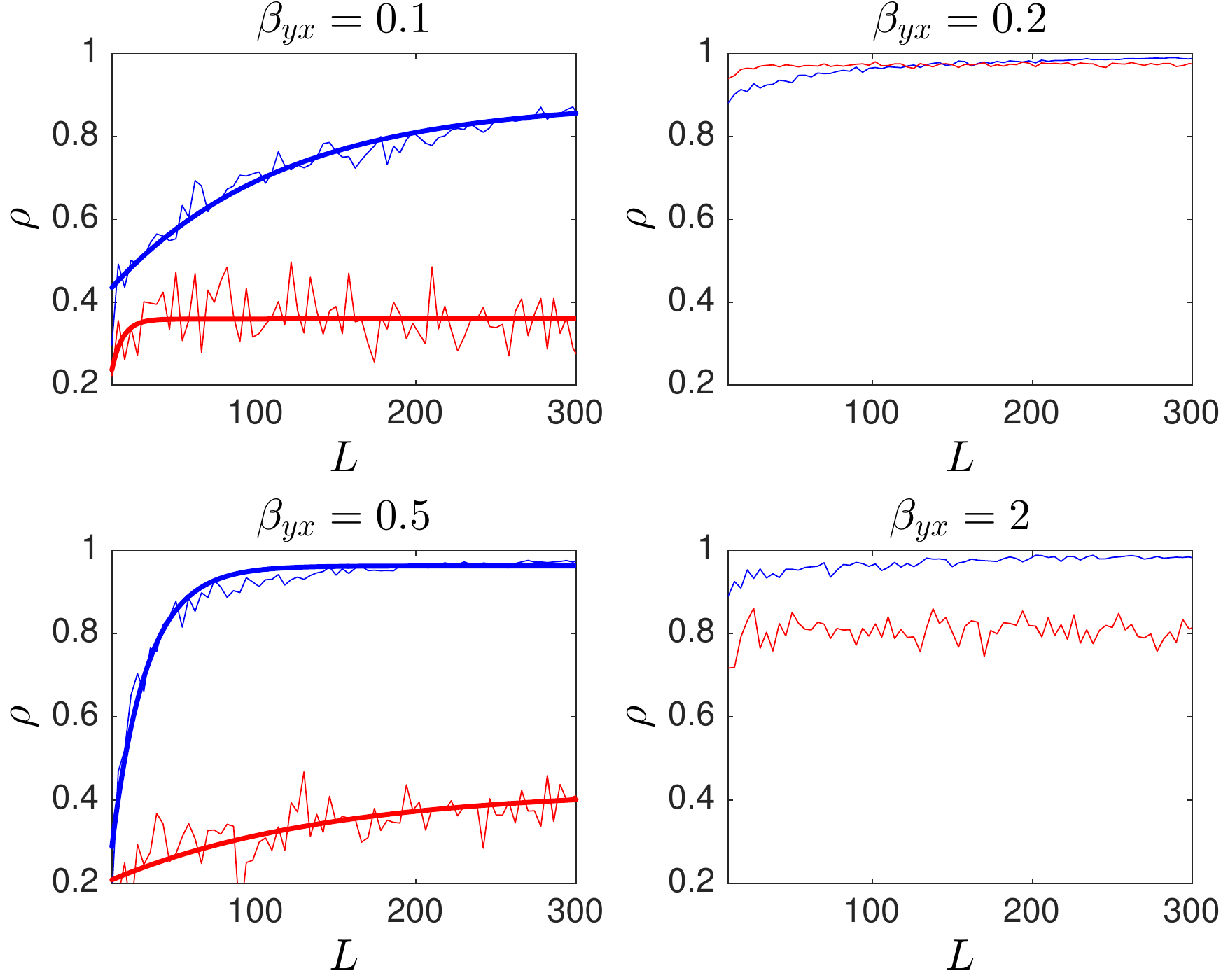}}
  \caption{Correlations between observed and cross-mapped estimates of $x$ (blue) and $y$ (red) for four different values of $\beta_{yx}$. Thin lines are the computed correlations and thick lines are fits to Eq.~\ref{eq:fit}. For $\beta_{yx}=0.2$ and $\beta_{yx}=2$ a good fit is not possible, and no fits are shown.} 
  \label{fig:beta}
  \vspace{-0.1cm}
\end{figure}

To study how CCM causality estimates depend on the strength of the coupling between the two variables we fix $r_x = 3.625$ and $r_y = 3.77$. To keep things simple, we will set $\beta_{xy}=0$ as in the previous example, and only vary $\beta_{yx}$. The correlation coefficients $\rho_{x\hat{x}}$ and $\rho_{y\hat{y}}$ are shown in Figure~\ref{fig:beta} for $\beta_{yx}=0.1, 0.2, 0.5, 2.0$. 

%
%

For $\beta_{yx} = 0.05$, a plot very similar to that displayed in Figure~\ref{fig:convergence} is produced (not shown in Figure~\ref{fig:beta}), which is consistent with a coupling from $x$ to $y$, but no coupling from $y$ to $x$. 

For $\beta_{yx} = 0.1$, $\rho_{x,\hat{x}}$ is seen to converge toward $\rho_\infty=0.89$, whereas $\rho_{y,\hat{y}}$ fluctuates around 0.36. This lack of convergence as a function of the library size is an indication that there is no coupling from $y$ to $x$.

For $\beta_{yx} = 0.5$, the convergence of $\rho_{x\hat{x}}$ is toward the higher value $\rho_\infty=0.96$, and convergence is faster, indicating a stronger coupling. Here we also see convergence of $\rho_{y\hat{y}}$ toward $\rho_\infty=0.43$, and although the value is much smaller than for $x$, this could indicate a coupling from $y$ to $x$. We know that this is not the case, since $\beta_{xy}=0$, so instead this is most likely due to the fact that $x$ is driving $y$, and that the two time series are becoming more synchronized.

For an intermediate $\beta_{yx}$ value of 0.2 something unexpected happens: both $\rho_{x\hat{x}}$ and $\rho_{y\hat{y}}$ quickly increase to values close to 1, already for very low values of $L$. Furthermore for $L \lesssim 100$, $\rho_{y\hat{y}} > \rho_{x\hat{x}}$. Naively this could be seen as an indication that there is a strong bidirectional coupling between $x$ and $y$. Alternatively both signals could be driven by a common external variable---the Moran effect \citep{moran_statistical_1953}---or one of the variables is driving the other so strongly that they synchronize \citep{boccaletti_synchronization_2002,pikovsky_synchronization:_2003}. We know, however, that neither the Moran effect, nor synchronization can be the explanation here, so a different explanation must be found. We return to this in section \ref{sec:bunching}, but note here that (i) for both $x$ and $y$ the correlation is uniformly high, i.e., there is not much evidence of convergence, and (ii) the function in Eq.~\ref{eq:fit} is not a good fit to the data.

For $\beta_{yx} = 2$ we again see very high correlation coefficients for both $x$ and $y$, but clearly $\rho_{x\hat{x}} > \rho_{y\hat{y}}$. There is not much evidence of convergence for $\rho_{x\hat{x}}$ and $\rho_{y\hat{y}}$ fluctuates around a value close to 0.8. This is consistent with synchronization due to the strong coupling from $x$ to $y$ and inspection of the time series confirms this. Again we note that $\rho(L)$ from Eq.~\ref{eq:fit} is not a good fit to the data.

\subsubsection{Near-periodicity and bunching\label{sec:bunching}}
\begin{figure}[!b]
  \centering
   {\includegraphics[width=\columnwidth]{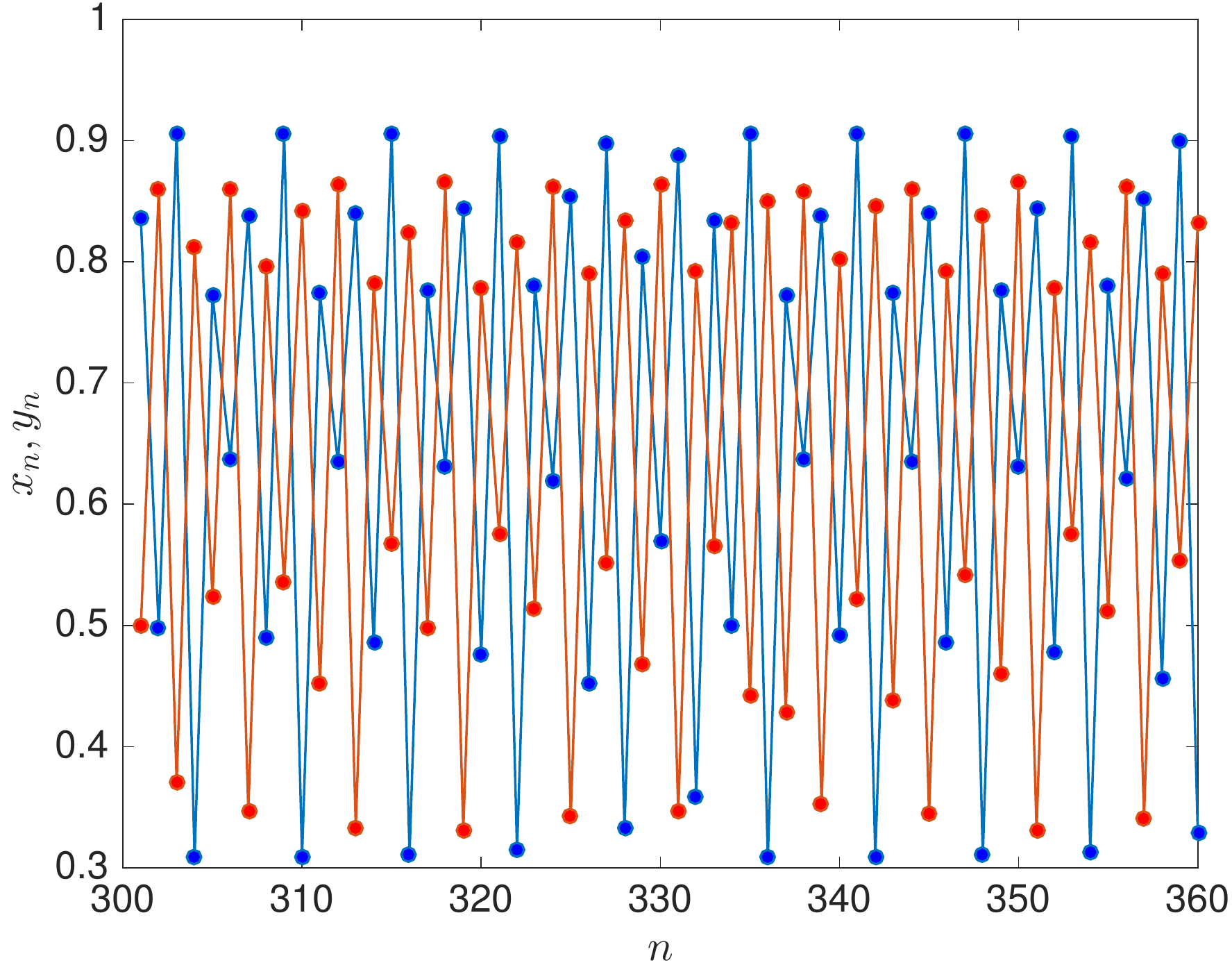}}
  \caption{Plot of the time series of $x$ (shown in blue) and $y$ (shown in red) displaying nearly periodic behavior. Parameter values: $r_x=3.625$, $r_y=3.77$, $\beta_{xy}=0$, $\beta_{yx}=0.2$.}
  \label{fig:bunching}
  \vspace{-0.1cm}
\end{figure}

We now return to the case in the upper right panel in Figure~\ref{fig:beta}. For $r_x=3.625$ and $r_y=3.77$ both logistic maps---considered in isolation---are in the chaotic phase, cf. Figure~\ref{fig:phase}. But for $\beta_{yx}=0.2$ the dynamics of $y$ become nearly periodic as illustrated in Figure~\ref{fig:bunching}.

\begin{figure}[!t]
  \centering
   {\includegraphics[width=\columnwidth]{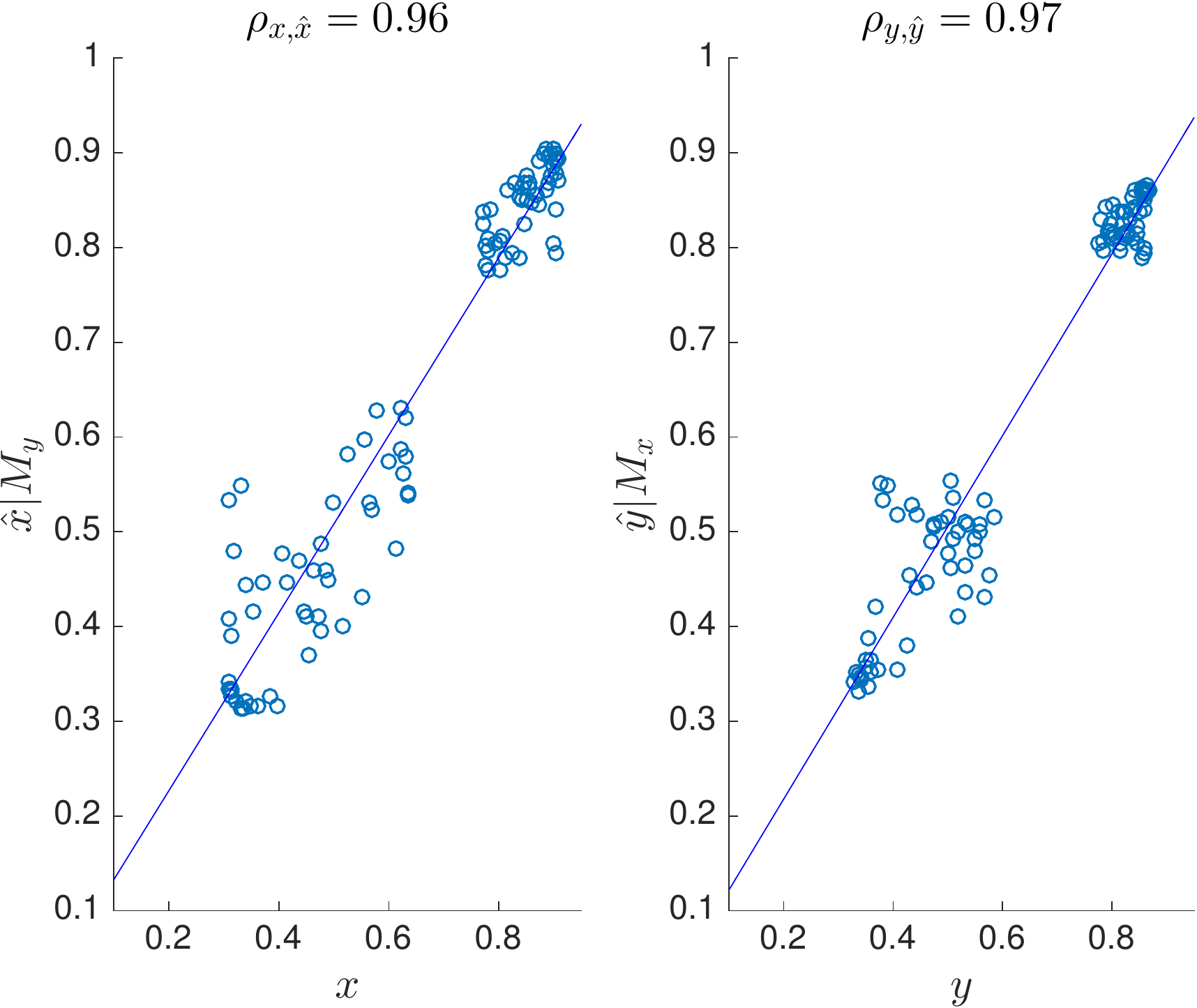}}
  \caption{Cross-mapped estimates vs.\ observed values of $x$ and $y$. Parameter values: $r_x=3.625$, $r_y=3.77$, $\beta_{xy}=0$, $\beta_{yx}=0.2$.}
  \label{fig:bunching_scatter}
  \vspace{-0.1cm}
\end{figure}

Nonlinear dynamic systems are known to display intermittency \citep{ott_chaos_2002}, where periods of chaos are interspersed with periods of nearly periodic behavior. Intermittency, however, does not seem to be the cause here, since the phenomenon persists no matter how many points are discarded as transient before starting to sample. Instead, it seems, that $y$ becomes partially synchronized with $x$ because the range of $x$-values lies within several relatively narrow intervals. An example of cross-mapped estimates for $x$ and $y$ are shown in figure~\ref{fig:bunching_scatter}.

Near-periodicity results in $\rho_{y\hat{y}}$ becoming higher than $\rho_{x\hat{x}}$ at low $L$, and `bunching' of the points leads to `fast non-convergence', i.e., high correlations, that are not reminiscent of the convergence behavior expected in CCM, evidenced by the fact that the fits to $\rho(L)$ in Eq.~\ref{eq:fit} fail.

\begin{figure}[!t]
  \centering
   {\includegraphics[width=\columnwidth]{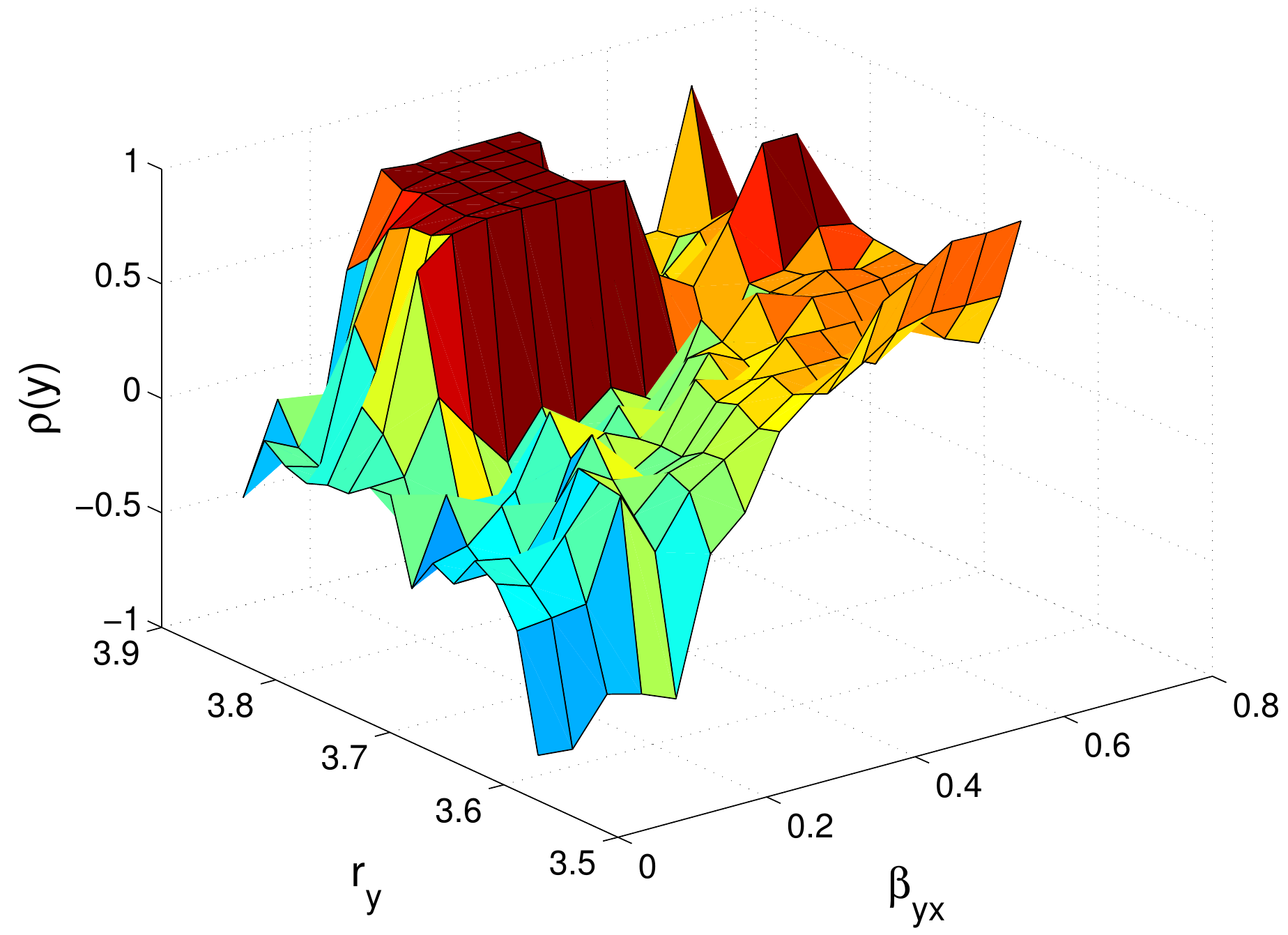}}
  \caption{Surface plot of $\rho_{y\hat{y}}$ for $L=400$ as a function of $r_y$ and $\beta_{yx}$. Note that $r_x$ is also varied, since $r_x+r_y=7.3984$. $\beta_{xy}=0$ throughout.}
  \label{fig:surf}
  \vspace{-0.1cm}
\end{figure}

To see whether this is an isolated point in parameter space, resulting in pathological behavior, we calculate cross-mapped estimates $\hat{y}\vert M_x$ of $y$ for a range of parameter values $r_x$, $r_y$ and $\beta_{yx}$. For each value we discard the first 1000 points in the time series and sample the next 400 points after that to construct the shadow manifolds. The result is shown in Figure~\ref{fig:surf}.

There is a clear plateau in Figure~\ref{fig:surf} that corresponds to the `bunching' observed in Figure~\ref{fig:bunching}, indicating that the issue is not confined to a few isolated parameter combinations. Whether similar issues affect the use of CCM for other models or when applied to empirical data is a topic for future research.

\subsection{Effects of noise}
Since all empirical data contain a certain level of noise it is important to study what effect  noise, as included in Eq.~\ref{eq:noise}, has on CCM results. We consider the simple case of unidirectional coupling from $x$ to $y$, i.e., $\beta_{xy}=0$. In this case the noise term $\epsilon_{x,t}$  propagates directly to the variable $y$, and therefore does not affect the cross-mapped estimates of $x$ based on $M_y$ (a fact that has been verified by numerical simulations). On the other hand, noise in $y$ is not coupled back into $x$ and could potentially be detrimental to the reconstruction. Hence, we will make the simplifying assumption of setting $\epsilon_{x,t}=0$.  This leaves $r_x, r_y, \beta_{yx}$ and the noise level $\sigma_y$ as free parameters of the model in Eq.~\ref{eq:noise}. We fix $r_x=3.8$ and $r_y=3.5$, so that $x$ is in the chaotic regime and the dynamics of $y$ is governed by a period-4 attractor for $\beta_{yx}=0$, but as illustrated in Figure~\ref{fig:phase} the dynamics of $y$ will become increasingly chaotic as $\beta_{yx}$ increases. We present results for three different values of $\beta_{yx}$ ranging from weak to strong coupling. At low noise levels CCM gives the correct result for all three values of the coupling strength, namely that $x$ causes $y$.

\begin{figure}[!h]
  \vspace{-0.2cm}
  \centering
   {\includegraphics[width=\columnwidth]{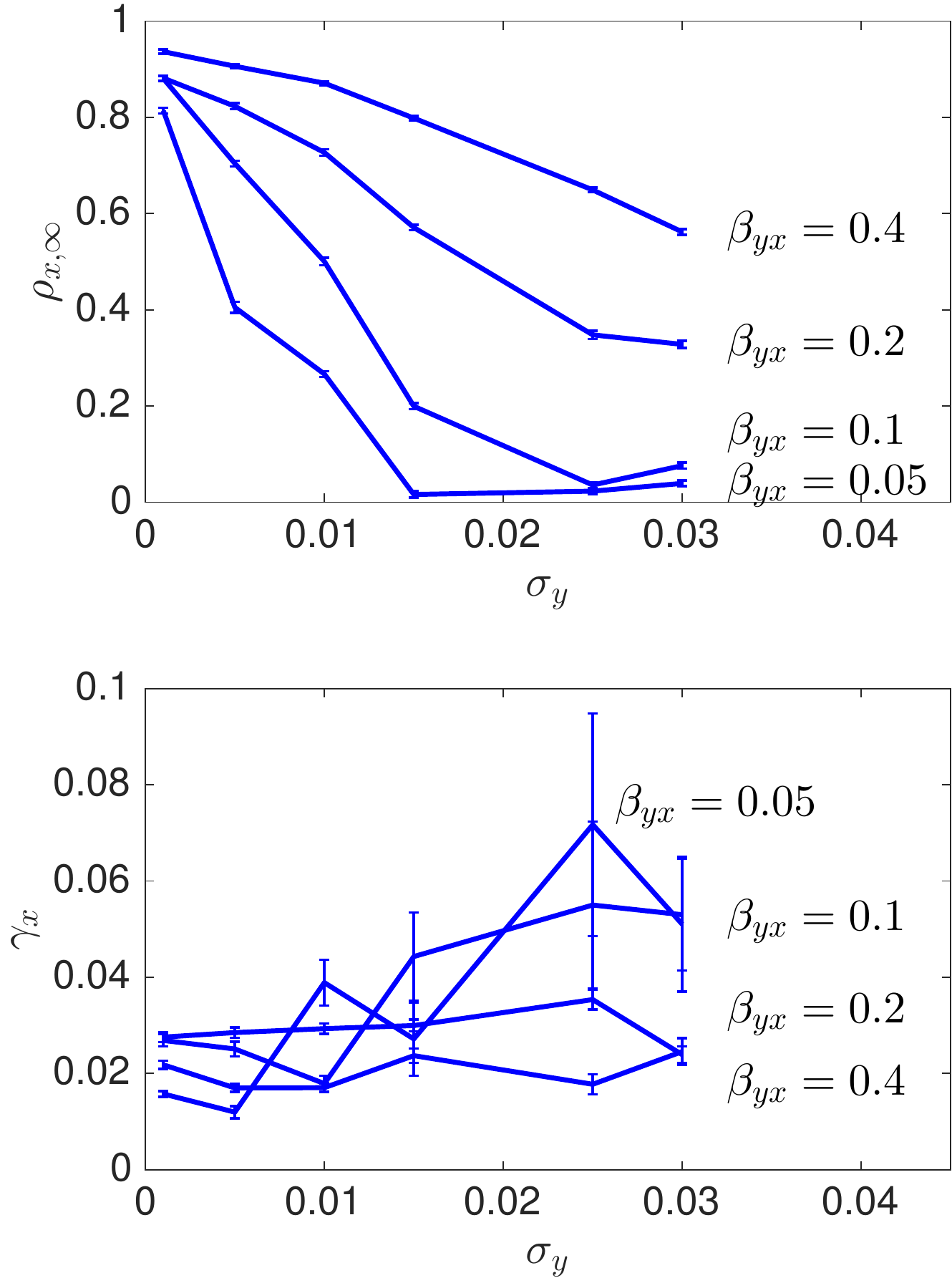}}
  \caption{The effect of the noise level $\sigma_y$ on the converged value $\rho_{x,\infty}$ of $\rho_{x\hat{x}}$ (top panel) and on the rate of convergence $\gamma_x$ of $\rho_{x\hat{x}}$ (bottom panel). In all cases $r_x=3.8$, $r_y=3.5$ and $\beta_{xy}=0$. The coupling $\beta_{yx}$ was varied as shown in the figure. The error bars represent 95\% confidence intervals on the fitted values.} 
  \label{fig:noise}
  \vspace{-0.1cm}
\end{figure}

When the noise level $\sigma_y$ is increased the cross-mapped estimates of $x$ from $M_y$ deteriorate, and as a result $\rho_{x\hat{x}}$ decreases.  Figure~\ref{fig:noise} shows the fitted values $\rho_{x,\infty}$ that $\rho_{x\hat{x}}$ converges to as a function of the noise level (top panel). 

We see that $\rho_{x,\infty}$  decreases in a roughly linear fashion as $\sigma_y$ is increased until the correlation becomes very low $(0\lesssim\rho_{x,\infty}\lesssim 0.2)$ where the curve becomes flat. For systems corresponding to the flat part of the curves $\rho_{x,\infty}$  is very low and at about the same level as $\rho_{y,\infty}$, so, effectively, the system is too noisy for CCM to extract the direction of causality from the converged value of $\rho_{x\hat{x}}$.

The bottom panel shows that the rate of convergence $\gamma_x$ from Eq.~\ref{eq:fit} is relatively unaffected by the noise level, except when $\rho_{x\hat{x}}\approx 0$, where CCM is not applicable under any circumstances. But for combinations of coupling strength and noise level that lie on the linear decline of the curves in the top panel of Figure~\ref{fig:noise} the fitted rate of convergence $\gamma_x$ seems to be a better indicator that $\rho_{x\hat{x}}$ converges than $\rho_{x,\infty}$; and hence also a better predictor of causality.

\subsubsection{Estimating coupling from noise}
\begin{figure}[!h]
  \vspace{-0.2cm}
  \centering
   {\includegraphics[width=\columnwidth]{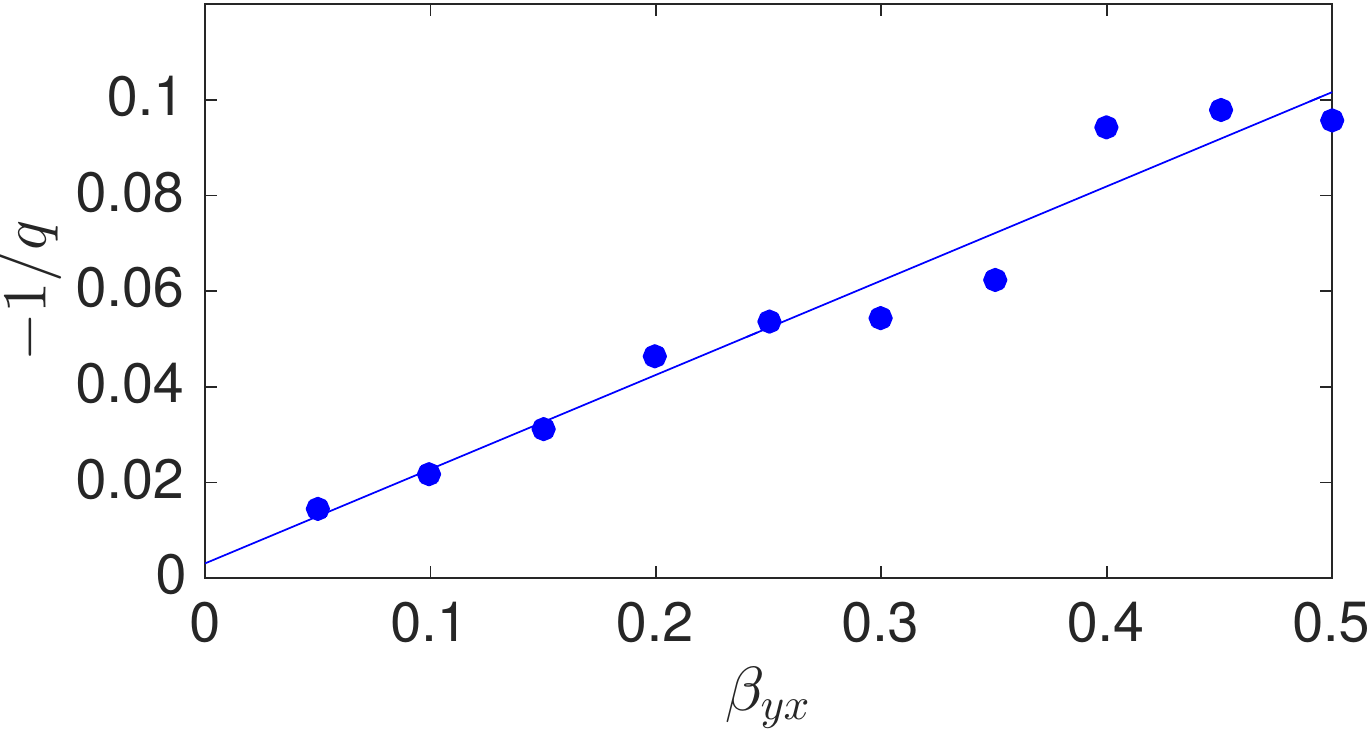}}
  \caption{The inverse negative slope $-1/q$ of the linear portion of the curves in the top panel of Figure~\ref{fig:noise} as a function of the coupling.} 
  \label{fig:slope}
  \vspace{-0.1cm}
\end{figure}

As noted above, we observe from Figure~\ref{fig:noise} that $\rho_{x,\infty}$ decreases linearly as a function of noise. We further note that the inverse of the slope depends linearly on the strength of the coupling, as shown in Figure~\ref{fig:slope}.
For the particular example studied here this has the implication that we can determine the strength of the coupling from $x$ to $y$ by controlled injection of noise into the system. Whether this results generalizes to other systems is a question for future research that has practical implications for measuring the interaction between different parts of complex dynamic systems.

%
%
\renewcommand{\floatpagefraction}{0.8}

\section{\uppercase{Discussion}}
\noindent Methods for reliable inference of causality between two or more variables are of great interest to many fields of research. Convergent Cross-Mapping (CCM) shows great potential in this regard. However, based on the systematic assessments presented in the previous sections, we conclude that the method can give erroneous results. It was noted already by \citet{sugihara_detecting_2012} that CCM does not correctly predict the direction of causality when the coupling is so strong that it results in synchronization of variables, as we have also seen above. Further, our study shows that the method seems very sensitive to the particular dynamics of the model system, and CCM also fails to correctly predict the direction of causality in cases where the coupling is weak to moderate. Interestingly, we have shown that both types of cases where CCM fails is associated with a failure to fit the observed values of cross-mapped correlations to the function $\rho(L)$ in Eq.~\ref{eq:fit}. It would therefore seem that the failure to produce a good fit to $\rho(L)$ is an indicator that CCM is not applicable to the data.

Another aspect of CCM performance concerns conditions of noise. Generally, we observe that CCM is fairly robust to random noise and makes reliable inferences at varying degrees of coupling under conditions of lower noise levels. However, for higher levels of noise CCM correlations drop linearly as a function of added noise. For systems subject to noise our results suggest that the rate of convergence is a more robust indicator of cross-mapping convergence and hence of causality. In applications where noise can be controlled, injecting noise into the system at different noise levels presents an opportunity to gauge the strength of coupling between variables.

Together our results warrant caution in the application of CCM to real-world data for purposes of causal inference, and care must be taken to look in detail at the convergence properties of the correlations between observed data and cross-mapped estimates. Our results suggest that fitting the correlation coefficients as a function of library length can give an indication of the applicability of CCM. When applied under the right circumstances, the method has the potential not only to inform the researcher about the causal direction of dynamics between coupled variables, but by controlled injection of noise, we can also infer the coupling strength between the variables.  
\section*{\uppercase{Acknowledgements}}
\noindent We would like to acknowledge the Interacting Minds Centre, Aarhus University, for providing the ideal environment for the authors' collaboration.

\vfill
\bibliographystyle{apalike}
{\small
\bibliography{CCM_Complexis_2016}}

\vfill
\end{document}